\documentclass[eqsecnum,showpacs,showkeys,floats,aps,nofootinbib,preprint]{revtex4}
\usepackage[latin9]{inputenc}
\usepackage{float}
\usepackage{amsmath}
\usepackage{amssymb}
\usepackage{graphicx}
\usepackage{esint}

\makeatletter
\@ifundefined{textcolor}{}
{%
 \definecolor{BLACK}{gray}{0}
 \definecolor{WHITE}{gray}{1}
 \definecolor{RED}{rgb}{1,0,0}
 \definecolor{GREEN}{rgb}{0,1,0}
 \definecolor{BLUE}{rgb}{0,0,1}
 \definecolor{CYAN}{cmyk}{1,0,0,0}
 \definecolor{MAGENTA}{cmyk}{0,1,0,0}
 \definecolor{YELLOW}{cmyk}{0,0,1,0}
}

\usepackage{bm}\usepackage{amsfonts}\usepackage{mathtools}\usepackage{color}\usepackage{wasysym}\usepackage{latexsym}\usepackage{epsfig}

\makeatother

\begin{document}
\makeatletter

\title{Bunching coefficients in Echo-Enabled Harmonic Generation}

\author{G. Dattoli}

\email{dattoli@frascati.enea.it}

\author{E. Sabia}

\email{sabia@frascati.enea.it}

\affiliation{ENEA - Centro Ricerche Frascati, via E. Fermi, 45, IT 00044 Frascati
(Roma), Italy}
\begin{abstract}
Coulombian diffusion determines a dilution of bunching coefficients
in Free Electron Laser seeded devices. From the mathematical point
of view the effect can be modeled through a heat type equation, which
can be merged with the ordinary Liouville equation, ruling the evolution
of the longitudinal phase space beam distribution. We will show that
the use of analytical tools like the Generalized Bessel Functions
and algebraic techniques for the solution of evolution problems may
provide a useful method of analysis and shine further light on the
physical aspects of the underlying mechanisms. 
\end{abstract}
\maketitle

\section{Introduction}

\label{s:intro} In this paper we will pursue some technical details
concerning the computation and the physical understanding of the recent
analysis by Stupakov \cite{Stupakov1} on the effect of Coulomb diffusion
on bunching coefficients in Echo Enabled Harmonic Generation (EEHG)
Free Electron Laser (FEL) seeded devices \cite{Stupakov2}. The considerations
developed in this paper should be understood as a complement to refs.
\cite{Stupakov1,Stupakov2}, with the aim of providing a more general
computational framework, benefitting from the formalism of beam transport
employing algebraic means \cite{Dattoli1}. In this treatment the
transport through magnetic elements is treated in terms of exponential
operator acting on an initial phase space distribution and usually
do not contain ``transport\textquotedblright{} elements provided
by a heat type diffusion. Here we will show that diffusion mechanisms
can be included in such a framework in a fairly straightforward way
\cite{Dragt}, preserving the spirit of exponential operator concatenation.

The effect of the Coulomb diffusion on the e-beam longitudinal phase
space distribution is ruled, according to \cite{Stupakov1}, by the
equation

\begin{eqnarray}
\partial_{s}f\left(p,\zeta,s\right)=D\partial_{p}^{2}f(p,\zeta,s),\label{eq:Oulomb diff}\\
f(p,\zeta,0)=f_{0}(p,\zeta)\nonumber 
\end{eqnarray}

Where $s$ is the coordinate of propagation, $\zeta$ is the longitudinal
coordinate in the beam, $p$ is associated to the beam energy, $D$
is the Coulomb diffusion coefficient and $f_{0}\left(p,\zeta\right)$
is the ``initial\textquotedblright{} distribution. We will discuss
the physical meaning of the previous quantities and we will provide
their explicit expression in the following, for the moment we note
that, having assumed that $D$ is independent of $p$ , we can solve
\eqref{eq:Oulomb diff} via the following Gauss-Weierstrass (G-W)
transform \cite{Babusci}

\begin{equation}
f\left(p,\zeta,s\right)=\frac{1}{2\sqrt{\pi Ds}}\int_{-\infty}^{\infty}e^{-\frac{\left(p-\eta\right)^{2}}{4Ds}}f_{0}(\eta,\zeta)d\eta\label{eq:GW}
\end{equation}

The bunching coefficients can be defined as those of the Fourier components
of the expansion \cite{Dattoli2}

\begin{equation}
f\left(p,\zeta,s\right)=\sum_{n=-\infty}^{+\infty}b_{n}\left(p,s\right)e^{in\zeta}\label{eq:Fourier components}
\end{equation}

and are, therefore, specified by 
\begin{equation}
b_{m}(p,s)=\frac{1}{2\pi}\int_{0}^{2\pi}d\zeta f\left(p,\zeta,s\right)e^{-im\zeta},
\end{equation}
The dilution of the bunching coefficients induced by the Coulomb diffusion
can therefore be mathematically expressed as

\begin{equation}
b_{m}(p,s)=\frac{1}{2\sqrt{\pi Ds}}\int_{-\infty}^{\infty}e^{-\frac{\left(p-\eta\right)^{2}}{4Ds}}b_{m}(\eta)d\eta\label{bunchingcoeffwith diffusion}
\end{equation}

Following ref. \cite{Stupakov1}, we assume the initial distribution

\begin{equation}
f_{0}(p,\zeta)=\frac{1}{\sqrt{2\pi}}e^{-\frac{\left[p-A_{1}\sin\left(\zeta-B_{1}p\right)\right]^{2}}{2}}\label{InitDistr}
\end{equation}

whose physical meaning will be discussed in the forthcoming section.
The solution of eq. \eqref{eq:Oulomb diff} can be therefore easily
computed, even though cannot be obtained in analytical form. In Fig.
1 we have reported the effect of the dispersion on the initial distribution
(Fig. 1a), after a drift $s=0.3$ with $D=0.7$ (Fig. 1b).

\begin{figure}[H]
$\qquad$\includegraphics[scale=0.38]{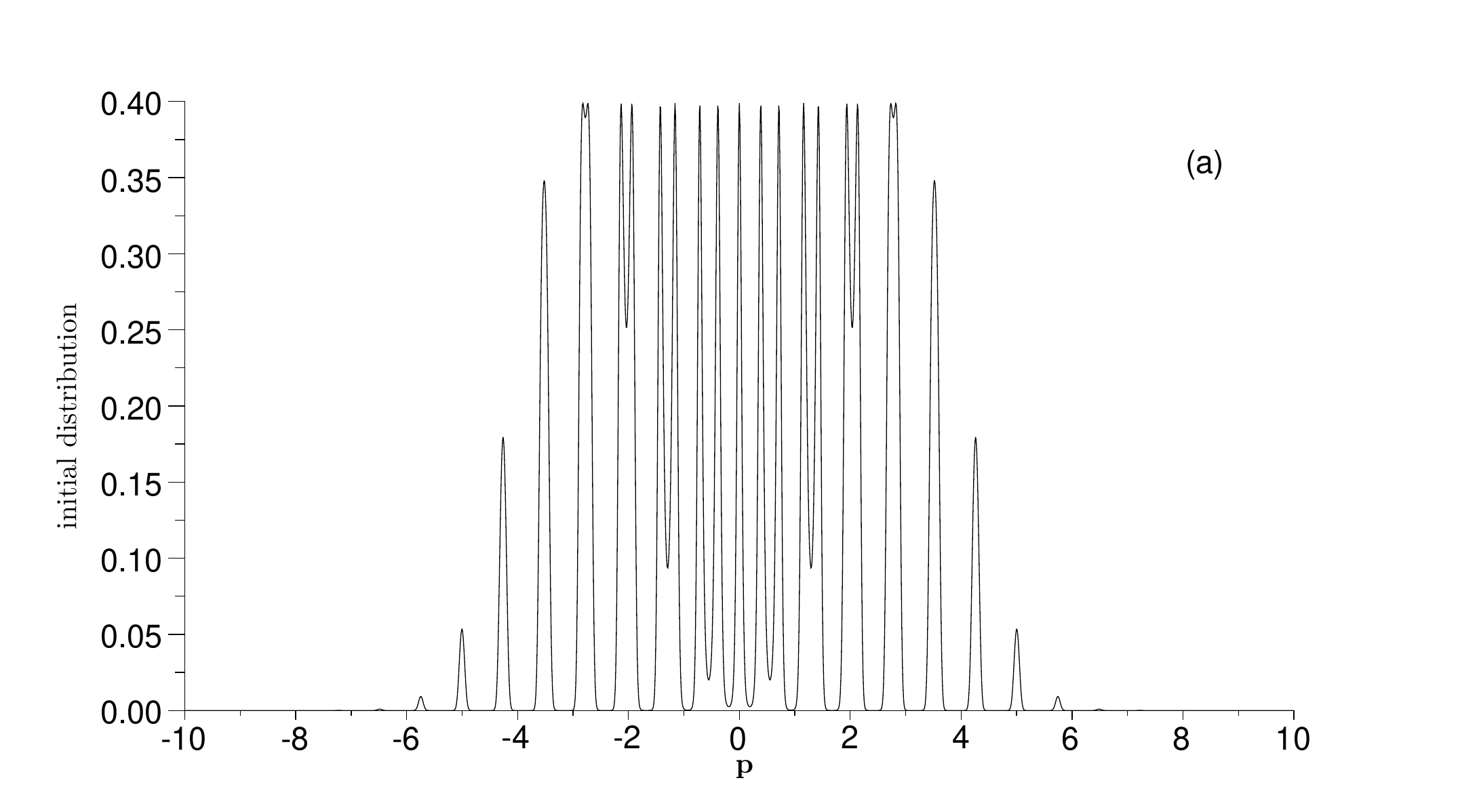}\includegraphics[scale=0.38]{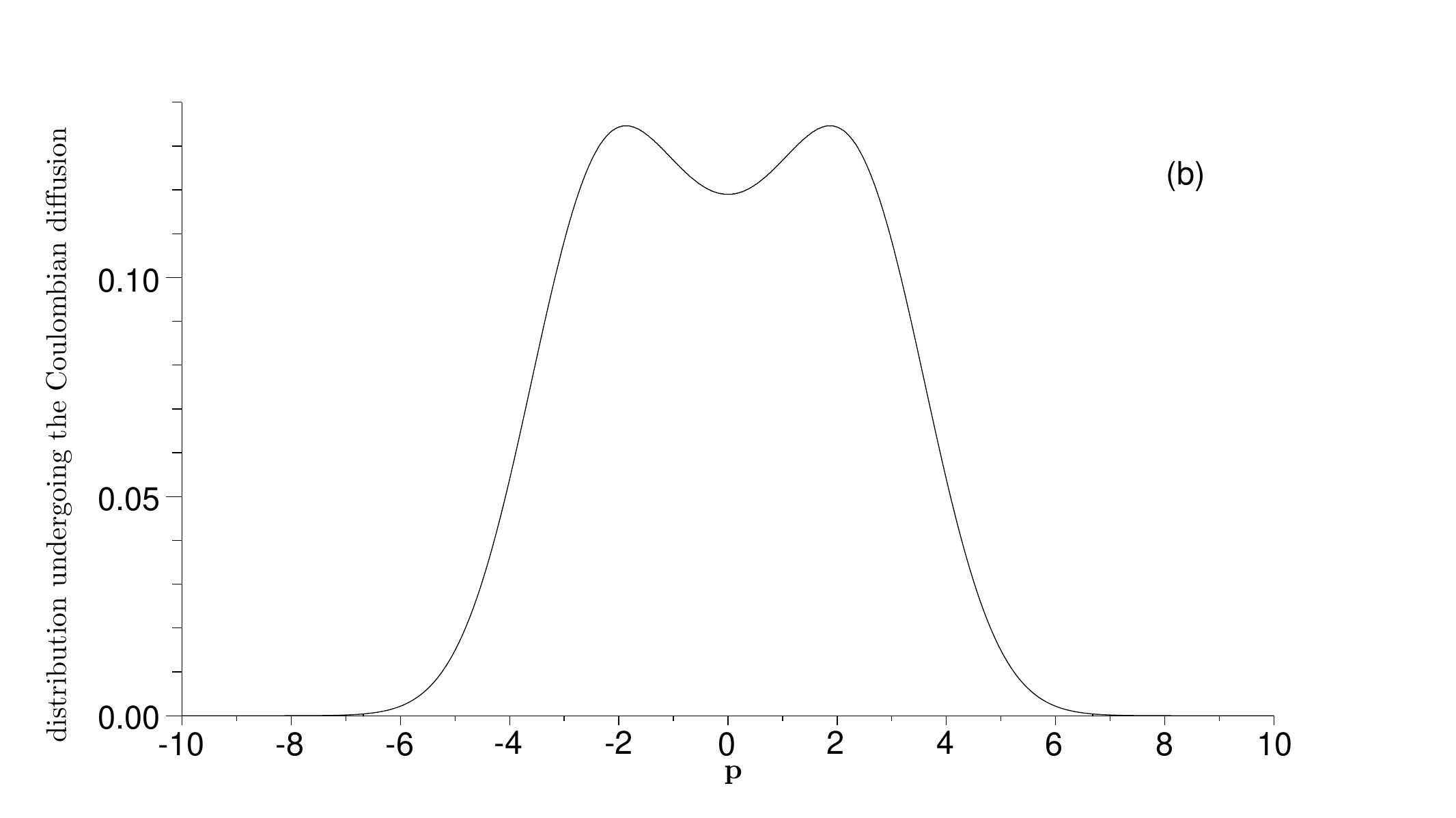}\caption{Evolution of the distribution function $f(p,\zeta,s)$ undergoing
the coulombian diffusion. a) $f(p,0,0),\: A_{1}=3,\: B_{1}=8.47$
b) $f(p,0,0.3),\: A_{1}=3,\: B_{1}=8.47,\: D=0.7$.}
\end{figure}

It is evident that the Coulomb term smears out the oscillations, associated
with the bunching coefficients, which tend to disappear or to be strongly
suppressed, as we will further discuss in the following. If we neglect
in eq. \eqref{InitDistr} the contributions $\left(A_{1}\sin\left(\zeta-B_{1}p\right)\right)^{2}$
we can expand $f_{0}(p,\zeta)$ in series of Bessel functions, namely

\begin{equation}
f_{0}(p,\zeta)\cong\frac{1}{\sqrt{2\pi}}e^{-\frac{p^{2}}{2}}e^{A_{1}p\sin\left(\zeta-B_{1}p\right)}=\frac{1}{\sqrt{2\pi}}e^{-\frac{p^{2}}{2}}\sum_{n=-\infty}^{\infty}e^{in\left(\zeta-B_{1}p\right)}J_{n}\left(-iA_{1}p\right)
\end{equation}
Where $J_{n}\left(\cdot\right)$ are cylindrical Bessel functions
of the first kind. The associated bunching coefficients are, neglecting
contributions in $A_{1}^{2}$ , provided by 
\begin{equation}
b_{m}(p)=\frac{1}{2\pi}\int_{0}^{2\pi}f_{0}\left(p,\zeta\right)e^{-im\zeta}d\zeta=\frac{1}{\sqrt{2\pi}}e^{-\frac{p^{2}}{2}}e^{-imB_{1}p}J_{m}\left(-iA_{1}p\right)\label{eqmot}
\end{equation}
The G-W transform can be accordingly written as (See Appendix) 
\begin{equation}
f\left(p,\zeta,s\right)\cong\frac{1}{2\sqrt{2\pi}\sqrt{1+2Ds}}e^{-\frac{p^{2}}{2\left(1+2Ds\right)}}\sum_{n=-\infty}^{\infty}e^{-\frac{\left(nB_{1}\right)^{2}Ds}{1+2Ds}}e^{\frac{inB_{1}p}{2\left(1+2Ds\right)}}e^{in\zeta}J_{n}\left(\frac{A_{1}\left(nB_{1}Ds-ip\right)}{1+2Ds}\right)\label{eq:GW Transorm}
\end{equation}
and bunching coefficients now read

$\,\,\,\,\qquad\qquad b_{m}(p,s)\cong e^{-\frac{\left(mB_{1}\right)^{2}Ds}{1+2Ds}}\Phi_{m}$

\begin{equation}
\Phi_{m}=\frac{1}{2\sqrt{2\pi}\sqrt{1+2Ds}}e^{-\frac{p^{2}}{2\left(1+2Ds\right)}}e^{\frac{imB_{1}p}{2\left(1+2Ds\right)}}J_{m}\left(\frac{A_{1}\left(mB_{1}Ds-ip\right)}{1+2Ds}\right)\label{eq:GW bunchimg}
\end{equation}

The effect of the Coulombian diffusion is therefore twofold, it induces

a) A dispersion due to the term $2Ds$

b) A suppression of the higher orders harmonics occurring through
$e^{-\frac{\left(mB_{1}\right)^{2}Ds}{1+2Ds}}$ .

In these introductory remarks we have provided a few remarks on the
type of formalism we are going to use to treat the bunching mechanism
in EEHG FEL seeded devices, the forthcoming sections will cover more
physical details.

\section{Liouville and Vlasov operators and bunching coefficient dynamics}

We have quoted the initial distribution given in eq. \eqref{InitDistr}
without any comment about its physical meaning. Its origin can be
traced back to the following Liouville equation 
\begin{equation}
\partial_{s}f(p,\zeta,s)=\hat{L}f(p,\zeta,s)\label{eq:Liouville equation}
\end{equation}

\[
\hat{L}=-Bp\partial_{\zeta}+AV'\left(\zeta\right)\partial_{p}
\]

\[
f_{0}\left(p,\zeta\right)=\frac{1}{\sqrt{2\pi}}e^{-\frac{p^{2}}{2}}
\]

which rules the evolution of an ensemble of non-interacting particles,
driven by the Hamiltonian 
\begin{equation}
H=B\frac{p^{2}}{2}+AV\left(\zeta\right)\label{Hamilton}
\end{equation}
The solution of eq. \eqref{eq:Liouville equation} can be cast in
the form 
\begin{equation}
f(p,\zeta,s)=\hat{U}(s)f_{0}\left(p,\zeta\right),
\end{equation}

\[
\hat{U}(s)=e^{\hat{L}s}
\]

We have denoted by $\hat{U}(s)$ and $\hat{L}$ the evolution and
Liouville operators, respectively, of our dynamical problem \cite{Dattoli1,Dragt}.
We note that the Liouville operator breaks into two non-commuting
parts, therefore any treatment of the associated evolution operator
demands for an approximate disentanglement of the exponential, which,
for simplicity will be assumed to be provided by \cite{Dattoli1,Dragt}
\begin{equation}
e^{\left(-Bp\partial_{\zeta}+AV'\left(\zeta\right)\partial_{p}\right)s}\cong e^{-Bsp\partial_{\zeta}}e^{AsV'\left(\zeta\right)\partial_{p}}
\end{equation}
which is accurate to $\frac{AB}{2}s^{2}\left[p\partial_{\zeta},V'(\zeta)\partial_{p}\right]$.

The use of the rule $e^{\lambda\partial_{x}}f(x)=f(x+\lambda)$ allows
to cast the solution of eq. \eqref{eq:Liouville equation} in the
form 
\begin{equation}
f(p,\zeta,s)\cong e^{-Bsp\partial_{\zeta}}e^{AsV'\left(\zeta\right)\partial_{p}}f_{0}(p,\zeta)=e^{-Bsp\partial_{\zeta}}f_{0}(p+AsV'(\zeta),\zeta)=f_{0}(p+AsV'(\zeta-Bsp),\zeta-Bsp).
\end{equation}
Therefore by setting 
\begin{equation}
V(\zeta)=\cos(\zeta),\, As=A_{1},\, Bs=B_{1}
\end{equation}
and keep as initial distribution a Gaussian in $p$ , we recover eq.
\eqref{InitDistr}.

The physical scheme, we are dealing with, is reported in ref. \cite{Stupakov2}),
to which the reader is addressed for further details. The e-beam undergoes
two successive modulations induced by two different lasers in two
different sections. The beam initial distribution \eqref{InitDistr}
is that occurring after the first chicane at the entrance of the second
modulator.

The physical content of the previous variables is the following

\begin{equation}
A_{1}=\frac{\Delta E_{1}}{\sigma_{E}},\, B_{1}=R_{5,6}\frac{k_{L}\sigma_{E}}{E_{0}},
\end{equation}

\[
p=\frac{E-E_{0}}{\sigma_{E}},\,\zeta=k_{L}z
\]

where $\sigma_{E}$ is the e-beam energy spread, $\Delta E_{1}$ is
the induced energy modulation, $k_{L}$ is the laser wave vector.

The dynamics of the bunching coefficients can be obtained from eqs.
\eqref{eq:Liouville equation} and \eqref{eq:Fourier components}
as ( $b_{n}(p,s)=b_{n}$) 
\begin{equation}
\partial_{s}b_{n}=-iBpnb_{n}+\frac{1}{2i}A\left[b_{n-1}-b_{n+1}\right].\label{Ziter}
\end{equation}

\[
b_{n}(p,0)=\delta_{n,0}e^{-\frac{p^{2}}{2}}
\]

Eq. \eqref{Ziter} yields an idea of the coupling between the various
bunching coefficients and has already been derived in a closely similar
fom in ref. \cite{Dattoli2}.

The inclusion of the Coulomb diffusion in this scheme can be modeled
by transforming the Liouville into a Vlasov equation, namely

\begin{equation}
\partial_{s}f(p,\zeta,s)=\left(D\partial_{p}^{2}+\hat{L}\right)f(p,\zeta,s)\label{Vlasov}
\end{equation}
The diffusion coefficient, expressed in practical units, reads \cite{Stupakov1}
\begin{equation}
D=1.55\frac{I\left[kA\right]}{\varepsilon_{x}\left[\mu m\right]\sigma_{x}\left[100\mu m\right]\left(\sigma_{E}\left[keV\right]\right)^{2}}
\end{equation}
with $\varepsilon_{x}$ and $\sigma_{x}$ being the emittance and
beam section respectively

The solution of eq. \eqref{Vlasov} can be obtained as

\begin{equation}
f(p,\zeta,s)\cong e^{Ds\partial_{p}^{2}}(e^{s\hat{L}}f_{0}(p,\zeta))
\end{equation}

According to this approximation the effect of the diffusion is calculated
separately from that induced by the Liouvillian contribution. Higher
order disentanglements \cite{Dattoli1,Dragt} can be used to get more
accurate results, but the present approximation is adequate for our
purposes.

In Fig. 2 we report the contour plots of the Liouville distribution
with and without the effect of the Coulombian diffusion for the same
s value. It is evident that the presence of a non-zero $D$ value
provides a significant reduction of the distribution harmonic content.
A global view is provided by Fig. 3 where we have reported the 3D
plot of the Liouville distribution under the action of diffusion for
different s-values.

\begin{figure}[H]
\includegraphics[scale=0.32]{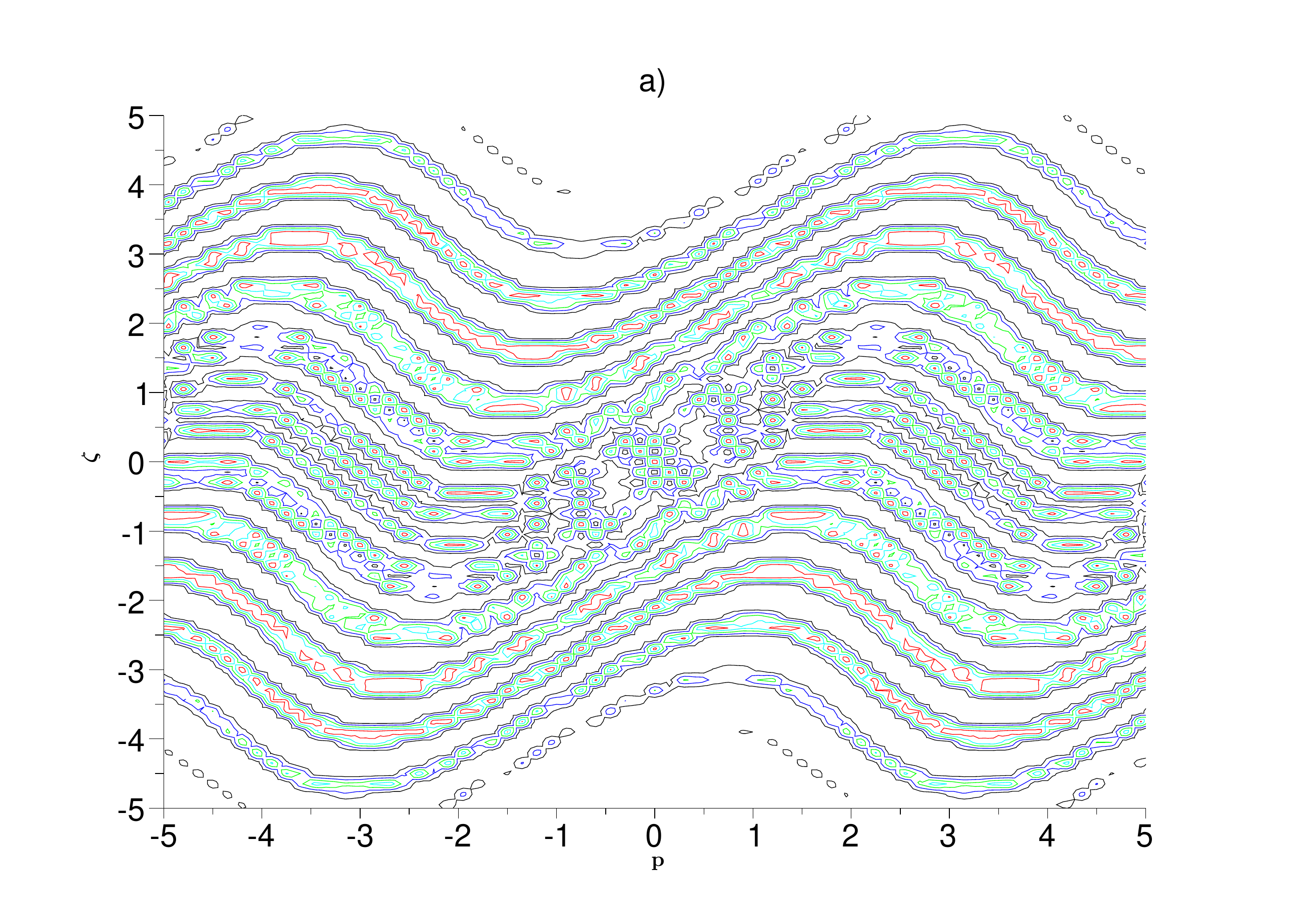}\includegraphics[scale=0.375]{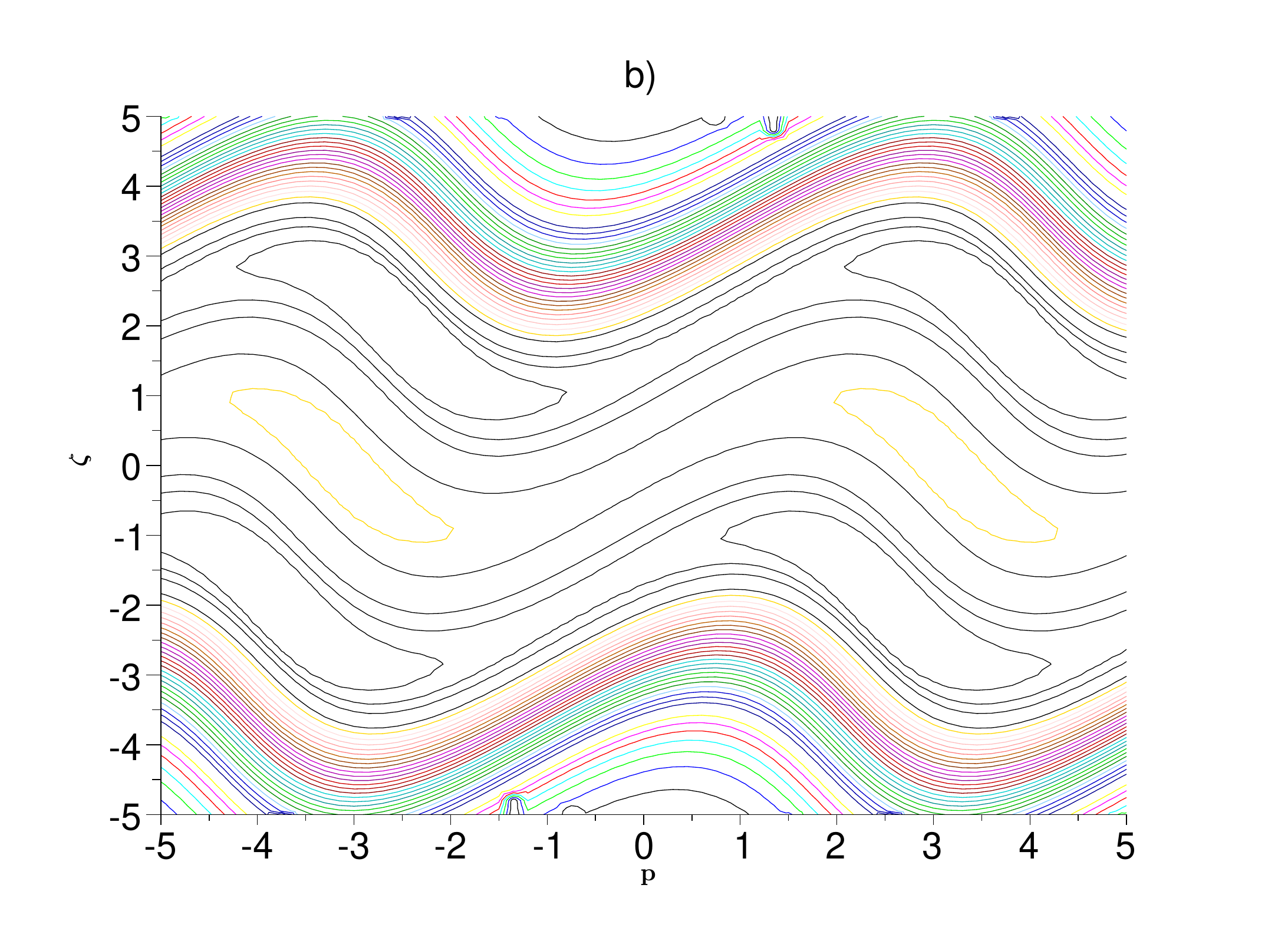}

\caption{Liouville distribution contour plots: a) without coulombian diffusion,
b) with coulombian diffusion $D=0.7$.}
\end{figure}

\begin{figure}[H]
\includegraphics[scale=0.3]{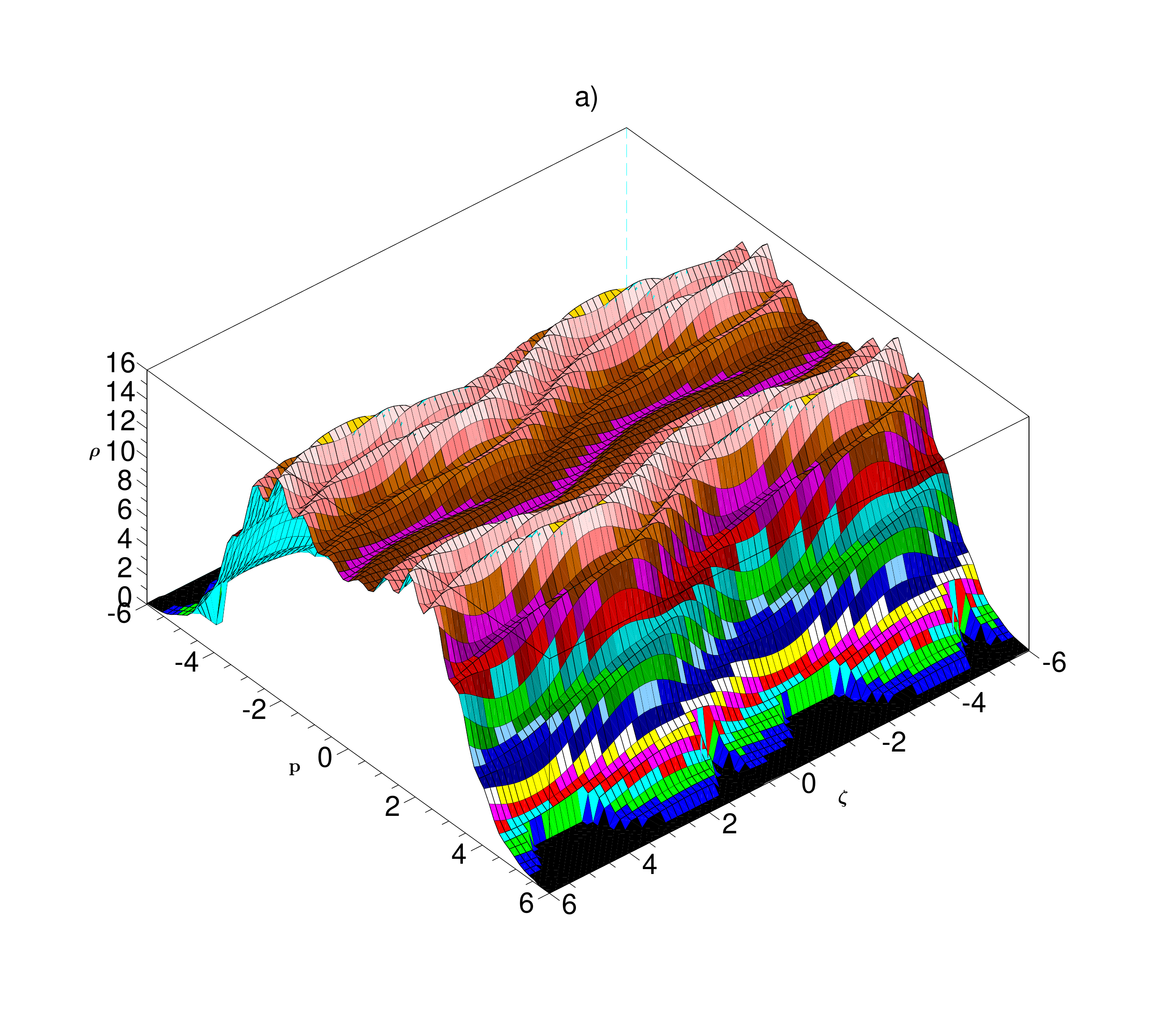}\includegraphics[scale=0.32]{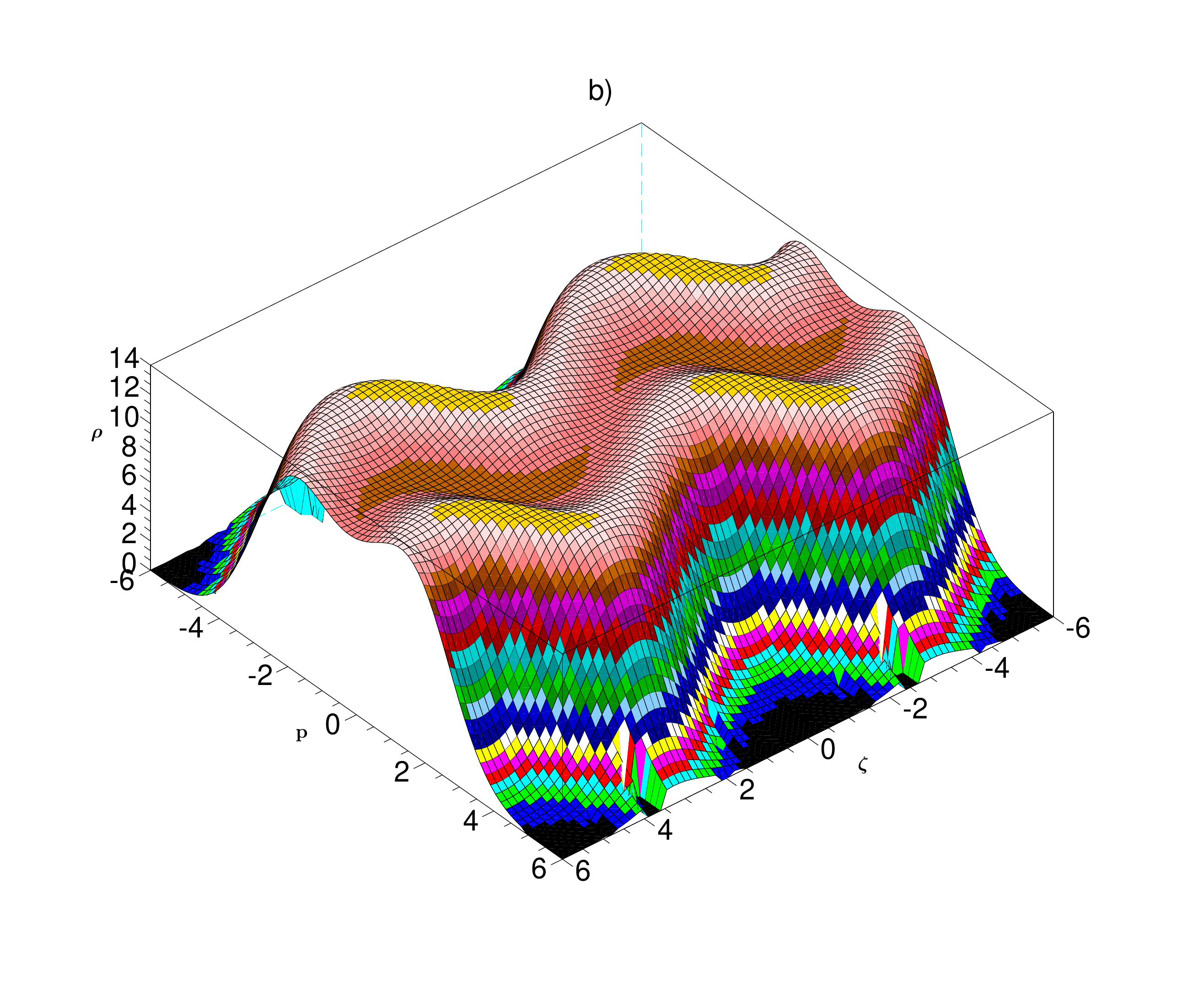}

$\qquad$$\qquad\qquad\qquad\qquad$\includegraphics[scale=0.35]{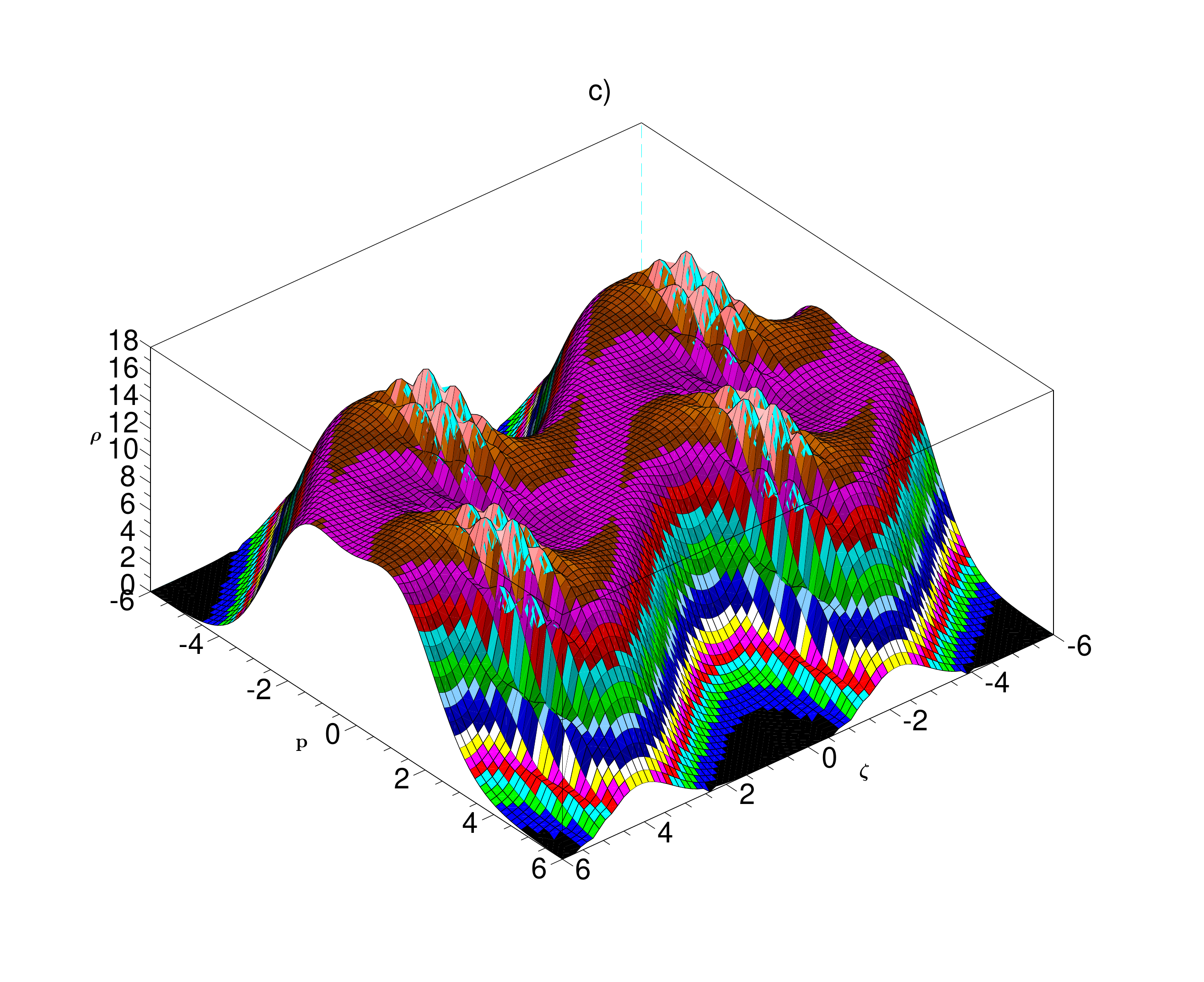}

\caption{Liouville distribution under the action of the coulombian diffusion
($D=0.7$) for different s values: a) $s=0.06$, b) $s=0.2$, c)$s=0.3$.}
\end{figure}

The nice feature of the procedure we are adopting is its modularity,
which will be further appreciated in the forthcoming section.

\section{Final Comments}

In the introductory section we have expanded the initial distribution,
by neglecting the contributions in $A^{2}$. Such an approximation
implies that the induced energy modulation is not large compared to
the natural e-beam energy spread. However if we relax these assumptions
and employ the formalism of generalized Bessel functions \cite{Babusci},
we can obtain a more general view on the problem under study. It has
indeed been shown that two variable Bessel functions can be expressed
through the generating function 
\begin{equation}
e^{ix\sin(\vartheta)-y(\sin(\vartheta))^{2}}=\sum_{n=-\infty}^{+\infty}e^{in\vartheta}\,_{H}J_{n}(x,y),\label{nleq}
\end{equation}

\[
\,_{H}J_{n}(x,y)=\sum_{r=0}^{\infty}\frac{\left(-1\right)^{r}}{2^{n+2r}}\frac{H_{n+2r}(x,y)}{r!(n+r)!},
\]

\[
H_{n}(x,y)=n!\sum_{r=0}^{n/2}\frac{x^{n-2r}y^{r}}{r!(n-2r)!}
\]

where $H_{n}(x,y)$ are two variable Hermite-Kamp� d� F�ri�t polynomials
\cite{Babusci}. They are actually understood as belonging to the
family of Hermite based functions and are widely exploited to deal
with radiation scattering problems \cite{Reiss}, going beyond the
dipole approximation.

It is also worth noting the following operational property \cite{Babusci}%
\footnote{According to the property \eqref{eq:two var bessel} the functions
$\,_{H}J_{n}(x,y)$ are solutions of the heat equation$\qquad$ $\qquad$$\partial_{y}F(x,y)=\partial_{x}^{2}F(x,y),$
$\:$$F(x,0)=J_{n}(x)$%
}

\begin{equation}
\,_{H}J_{n}(x,y)=e^{y\partial_{x}^{2}}J_{n}(x)\label{eq:two var bessel}
\end{equation}
which will be used in the following.

The inclusion of the generalized Bessel functions does not change
substantively the form of the bunching coefficient, which are obtained
from eqs. \eqref{eq:GW Transorm} and \eqref{eq:GW bunchimg} by replacing
the Bessel function with

\[
\,_{H}J_{n}(\frac{A_{1}(nB_{1}Ds-ip)}{1+2Ds},\frac{A_{1}^{2}}{1+2Ds})
\]

In the second modulator the initial distribution will be provided
by the properly modified eq. \eqref{eq:GW Transorm} so that

\begin{equation}
f_{2}(p,\zeta)\cong e^{-B_{2}p\partial_{\zeta}}e^{-A_{2}\sin(\zeta)\partial_{p}}f_{1}(p,\zeta)
\end{equation}
which holds if the laser, in the second modulator, is the same of
the first. We have removed the $s$ variable because absorbed into
the $A$ and $B$ terms. If we use the Bessel function expansion 
\begin{equation}
e^{-A_{2}\sin(\zeta)\partial_{p}}=\sum_{n=-\infty}^{\infty}e^{in\zeta}J_{n}(iA_{2}\partial_{p})
\end{equation}
we can therefore cast the distribution at the end of the second modulator
as 
\begin{equation}
f_{2}(p,\zeta)\cong\sum_{n=-\infty}^{\infty}e^{in(\zeta-B_{2}p)}J_{n}(iA_{2}\partial_{p})f_{1}(p,\zeta-B_{2}p)=
\end{equation}

\[
=\frac{1}{\sqrt{2\pi}}\sum_{n=-\infty}^{\infty}e^{in(\zeta-B_{2}p)}J_{n}(iA_{2}\partial_{p})e^{-\frac{p^{2}}{2}}\sum_{m=-\infty}^{\infty}e^{im(\zeta-(B_{1}-B_{2})p)}J_{m}(-iA_{1}p)
\]

By rearranging the indices we end up with 
\begin{equation}
f_{2}(p,\zeta)\cong\frac{1}{\sqrt{2\pi}}\sum_{l=-\infty}^{\infty}e^{il\zeta}Q_{l}\label{eq:distrib. with bessel}
\end{equation}

\[
Q_{l}=\sum_{r=0}^{l}e^{-i(l-r)B_{2}p}J_{l-r}(iA_{2}\partial_{p})e^{-\frac{p^{2}}{2}}e^{-i(l-r)(B_{1}+B_{2})p}J_{r}(-iA_{1}p)
\]

According to such a picture the n-th bunching coefficient is characterized
by a discrete convolution of the bunching terms of the second modulator
on the first.

The effect of the Coulombian diffusion on the distribution \eqref{eq:distrib. with bessel}
can be easily computed, either numerically or analytically. By keeping
the lowest order contribution $l=1$ and by approximating the Bessel
function as

\[
J_{0}(x)\cong1,\: J_{1}(x)\cong\frac{x}{2}
\]
we obtain 
\begin{equation}
Q_{1}\cong-\left[\left(\frac{iA_{1}p}{2}\right)+(iA_{2}p)e^{-i(B_{1}+2B_{2})p}-\frac{B_{1}+B_{2}}{2}A_{2}e^{-i(B_{1}+2B_{2})p}\right]e^{-\frac{p^{2}}{2}}.\label{cont}
\end{equation}

From the above equation we easily deduce that the Coulombian contribution
yields a suppression factor proportional to $e^{-\frac{(B_{1}+2B_{2})^{2}Ds}{1+4Ds}}$
. More in general as also pointed in ref. \cite{Stupakov1} the larger
is the order of bunching the more significant is the effect of the
reduction.

The bunching coefficients vs. s in the presence of diffusion is provided
by Fig. 4.

\begin{figure}[H]
$\qquad\qquad\quad\qquad\qquad\qquad$\includegraphics[scale=0.5]{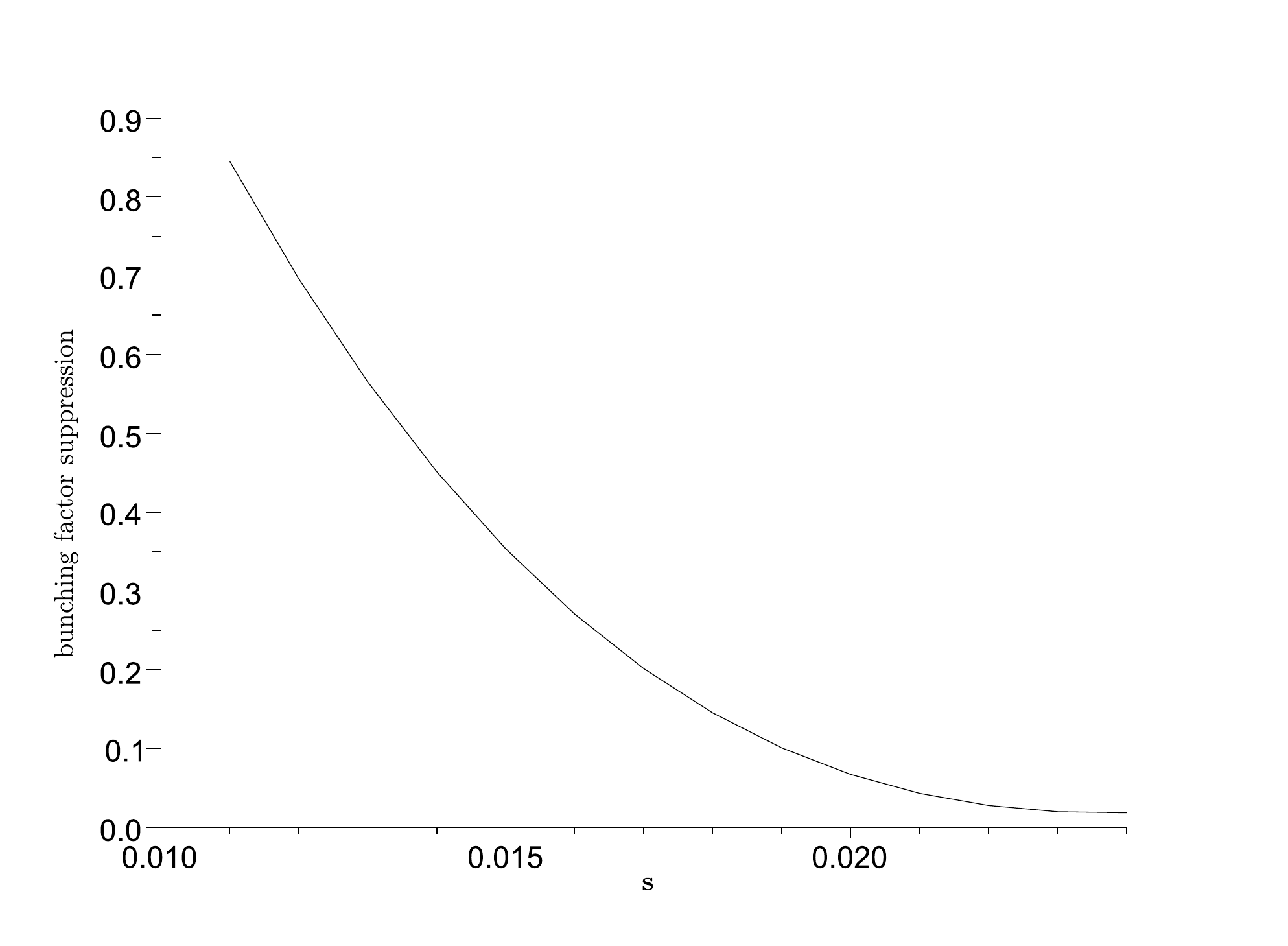}\caption{Effect of the coulombian diffusion ( $D=0.7$ ) on the bunching coefficient
( $m=50$) $b_{50}(D)/b_{50}(0)$ vs. s}
\end{figure}

The formalism we have presented in this paper provides a fairly detailed
analysis of how the Coulombian diffusion affect the bunching coefficients
in Echo enabled FEL devices. Our treatment is complementary to the
seminal contributions of refs. \cite{Stupakov1,Stupakov2} and confirm,
within a more general framework, their conclusions. We believe that
the modularity of the methods we have described in this paper offer
further opportunities as e. g. that of including in the treatment
more complicated arrangements of transport lines and of external fields.
The possibility of embedding these procedures with diffusion and damping
in storage rings, will be discussed elsewhere.

\begin{center}
\textbf{\large Appendix}{\large {} } 
\par\end{center}

We will discuss here some technical details concerning the solution
of the heat equation

\[
\partial_{t}F(x,t)=\partial_{x}^{2}F(x,t),
\]

\[
F(x,0)=x^{n}e^{-x^{2}}
\]

which can formally be written as 
\[
F(x,t)=e^{t\partial_{x}^{2}}(x^{n}e^{-x^{2}})
\]
The use of operational methods \cite{Babusci} yields 
\[
F(x,t)=(e^{t\partial_{x}^{2}}x^{n}e^{-t\partial_{x}^{2}})e^{t\partial_{x}^{2}}e^{-x^{2}}=
\]

\[
=(x+2t\partial_{x})^{n}e^{t\partial_{x}^{2}}e^{-x^{2}}.
\]

Then by applying the Glaisher and Burchnall \cite{Babusci} rules,
respectively reported below 
\[
e^{t\partial_{x}^{2}}e^{-x^{2}}=\frac{1}{\sqrt{1+4t}}e^{-\frac{x^{2}}{1+4t}},
\]

\[
(x+2t\partial_{x})^{n}=\sum_{s=0}^{n}\left(\begin{array}{c}
n\\
s
\end{array}\right)H_{n-s}(x,t)(2t\partial_{x})^{s}
\]

provides the solution of our problem in the form 
\[
F(x,t)=\frac{1}{\sqrt{1+4t}}\sum_{s=0}^{n}\left(\begin{array}{c}
n\\
s
\end{array}\right)H_{n-s}(x,t)(2t\partial_{x})^{s}e^{-\frac{x^{2}}{1+4t}}.
\]

The identities \cite{Babusci} 
\[
\partial_{x}^{s}e^{-\alpha x^{2}}=(-1)^{s}H_{s}(-2\alpha x,-\alpha),
\]

\[
H_{n}(\alpha_{1}+\alpha_{2},\beta_{1}+\beta_{2})=\sum_{s=0}^{n}\left(\begin{array}{c}
n\\
s
\end{array}\right)H_{n-s}(\alpha_{1},\beta_{1})H_{s}(\alpha_{2},\beta_{2})
\]

finally provide the result 
\[
F(x,t)=\frac{1}{\sqrt{1+4t}}\sum_{s=0}^{n}\left(\begin{array}{c}
n\\
s
\end{array}\right)H_{n-s}(x,t)H_{s}(-\frac{4xt}{1+4t},-\frac{4t^{2}}{1+4t})e^{-\frac{x^{2}}{1+4t}}=
\]

\[
=\frac{1}{\sqrt{1+4t}}H_{n}(\frac{x}{1+4t},\frac{t}{1+4t})e^{-\frac{x^{2}}{1+4t}}.
\]

It is accordingly evident that for any initial condition such that
\[
F(x,0)=g(x)e^{-x^{2}}
\]

with

\[
g(x)=\sum_{p}a_{p}x^{p}
\]

\[
F(x,t)=\frac{1}{\sqrt{1+4t}}\sum_{p}a_{p}H_{p}(\frac{x}{1+4t},\frac{t}{1+4t})e^{-\frac{x^{2}}{1+4t}}.
\]


\end{document}